\documentclass[a4paper,11pt]{article}

\pdfoutput=1 

\usepackage{jcappub} 

\usepackage[T1]{fontenc} 
\usepackage{siunitx}[separate-uncertainty = true, multi-part-units=single]
\usepackage{caption}
\usepackage{subcaption}
\usepackage{multirow}
\usepackage{array}
\newcolumntype{C}[1]{>{\centering\arraybackslash}p{#1}}

\title{$\texttt{SwiftC}_\ell$: fast differentiable angular power spectra beyond Limber}

\author[a]{Laura Reymond,}
\author[a]{Alexander Reeves,}
\author[a, b, c]{Pierre Zhang,}
\author[a]{Alexandre Refregier}

\affiliation[a]{ETH Zürich, Institute for Particle Physics and Astrophysics, Wolfgang-Pauli-Strasse 27, CH-8093 Zürich, Switzerland}
\affiliation[b]{Institute for Theoretical Physics, ETH Zürich, 8093 Zürich, Switzerland}
\affiliation[c]{Dipartimento di Fisica “Aldo Pontremoli”, Università degli Studi di Milano, 20133 Milan, Italy}

\emailAdd{lreymond@phys.ethz.ch}

\abstract{The upcoming stage IV wide-field surveys will provide high precision measurements of the large-scale structure (LSS) of the universe. Their interpretation requires fast and accurate theoretical predictions including large scales. For this purpose, we introduce $\texttt{SwiftC}_\ell$, a fast, accurate and differentiable $\texttt{JAX}$-based pipeline for the computation of the angular power spectrum beyond the Limber approximation. It uses a new FFTLog-based method which can reach arbitrary precision and includes interpolation along $k$, allowing for $k$-dependent growth factor and biases. $\texttt{SwiftC}_\ell$ includes a wide range of probes and effects such as galaxy clustering, including magnification bias, redshift-space distortions and primordial non-Gaussianity, weak lensing, including intrinsic alignment, cosmic microwave background (CMB) lensing and CMB integrated Sachs-Wolfe effect. We compare our pipeline to the other available beyond-Limber codes within the N5K challenge from the Rubin Observatory Legacy Survey of Space and Time (LSST) Dark Energy Science Collaboration. $\texttt{SwiftC}_\ell$ computes the 120 different angular power spectra over 103 $\ell$-multipoles in 5 ms on one GPU core while the computation of the gradient is approximately 4$\times$ slower. Using a pre-calculation, $\texttt{SwiftC}_\ell$ is thus about 40$\times$ faster than the winner of the N5K challenge with comparable accuracy. Furthermore, all outputs are auto-differentiable, facilitating gradient-based sampling and robust and accurate Fisher forecasts. We showcase a Markov Chain Monte Carlo, a Hamiltonian Monte Carlo and a Fisher forecast on an LSST-like survey, illustrating $\texttt{SwiftC}_\ell$'s differentiability, speed and reliability in measuring cosmological parameters. The code is publicly available at \url{https://cosmo-gitlab.phys.ethz.ch/cosmo_public/swiftcl}.}

\begin{document}
\maketitle
\flushbottom

\section{Introduction}
\label{sec:Introduction}
With the upcoming data releases from stage IV large-scale structure surveys such as the Vera C. Rubin Observatory's Legacy Survey of Space and Time (LSST) \cite{LSST}, Euclid \cite{Euclid}, the Dark Energy Spectroscopic Instrument (DESI) \cite{DESI} and SphereX \cite{spherex} arises the need for more accurate and numerically efficient analysis pipelines. Furthermore, current optical surveys such as the Dark Energy Survey (DES) \cite{DES} and Kilo-Degree Survey (KiDS) \cite{KiDS} and spectroscopic surveys such as the Baryonic Oscillation Spectroscopic Survey (BOSS) \cite{BOSS} already provide a vast amount of data available for analysis. The challenges for the analysis pipelines are twofold: the high precision of the data calls for accurate pipeline analysis, while the unprecedented amount of data available requires efficient pipelines. 

The redshifts of objects observed in these surveys are sometimes known only approximately through an average line-of-sight distribution as it is the case when targeted through photometry. The key observable is then the projected Fourier two-point correlation along the line-of-sight, namely, the projected angular power spectrum. Because of the projection, the computation of the angular power spectrum remains a numerically complex problem due to the presence of the Bessel function, a highly oscillatory function, in the integrand. For this reason, until recently the computation was mostly performed using the Limber approximation \cite{Limber} or the extended Limber approximation \cite{Limber++}. While this approximation allows for a great simplification of the computation and satisfies the required accuracy at small angular scales, this approximation breaks down at larger angular scales.  Some effort has been made in the community to develop more accurate methods of computation as can be seen for example in \texttt{AngPow} \cite{Angpow} or in the N5K challenge \cite{N5KChallenge} for the LSST survey. The winner of the challenge, \texttt{FKEM} \cite{Fang_2020}, uses a FFTLog decomposition of the window function and allows for a fast and accurate evaluation of the angular power spectrum. Furthermore, \texttt{BLAST.jl} \cite{blast}, a \texttt{Julia}-based code that relies on a decomposition of the 3D power spectrum into Chebyshev polynomials, was developed with a similar objective in mind. Some effort has also been made in the development of differentiable codes using the \texttt{Python}-$\texttt{JAX}$ library \cite{JAX} such as \texttt{JAX-COSMO} \cite{jax-cosmo} or similarly \texttt{Limberjack.jl} \cite{Limberjack} written in \texttt{Julia}. It is however worth noting that these pipelines use the Limber approximation.

When using the Limber approximation, large scales need to be removed from the analysis due to its inaccuracy in this regime. However, some effects such as primordial non-Gaussianity (PNG) or CMB integrated Sachs-Wolfe (ISW) effect can only be observed at these scales. In these cases, accurate theoretical predictions at large scales become crucial. Furthermore, ignoring the error introduced by the Limber approximation can lead to biases in the size and correlations of parameter confidence intervals in Fisher forecasts in the context of upcoming wide-angle surveys such as Euclid \cite{Bellomo_2020}.

It is for these reasons that we developed $\texttt{SwiftC}_\ell$, a fast, accurate and differentiable beyond-Limber pipeline for the computation of the angular power spectrum. It is based on a new FFTLog-based method that uses interpolation along the $k$-range, fully capturing the $k$-dependence and redshift dependence of the power spectrum, even, and especially, when the two cannot be factorised. We include a wide range of probes and effects needed for the analyses of stage IV surveys. The $3\times2$pt set-up can be easily done in $\texttt{SwiftC}_\ell$ with implemented functions for galaxy clustering, including magnification bias, redshift-space distortions and local primordial non-Gaussianity and weak lensing, including intrinsic alignment. In addition to these probes, we also include the CMB lensing and CMB ISW effect. Furthermore, the use of the \texttt{Python}-$\texttt{JAX}$ library allows for automatic differentiation at all steps of the calculation and thus makes $\texttt{SwiftC}_\ell$ ideal for gradient-based sampling and Fisher forecasts.

This paper is structured as follows: we first start by setting up the necessary theoretical framework in section $\S$\ref{sec:Theory}. Then in section $\S$\ref{sec:angular power spectrum}, we detail our FFTLog method for the computation of the angular power spectrum and show the Limber approximation. In section $\S$\ref{sec:Results}, after assessing internally the accuracy and performance of the pipeline, we compare $\texttt{SwiftC}_\ell$ to various existing libraries and codes. A Markov Chain Monte Carlo (MCMC), a Hamiltonian Monte Carlo (HMC) and a Fisher analysis on LSST-like data are also showcased for illustration. Finally, we conclude in section $\S$\ref{sec:Conclusion}. Numerical accuracy convergence tests on $\texttt{SwiftC}_\ell$ are provided in Appendix~\ref{appendix}.

\vfill
\section{Theory}
\label{sec:Theory}

\subsection{Definition of the angular power spectrum}
The angular power spectrum at multipole $\ell$ between two cosmological probes labelled by $i,j$ can be defined as follows \cite{Dodelson}:
\begin{equation}
    C_\ell^{ij} = \frac{2}{\pi}\int \text{d}k \ k^2 \Delta_\ell^i(k)\Delta_\ell^j(k) P(k,\chi,\chi'),
\end{equation}
with $\chi$ and $\chi'$ the comoving distances, $k$ the wavenumber and $P$ the matter power spectrum at unequal time. $\Delta_\ell^{i,j}$ are the line-of-sight kernels with $i,j$ the available probes in $\texttt{SwiftC}_\ell$. In this work, we consider the following cosmological probes: galaxy clustering contributions are written as $i,j = (g, D), (g,\mu), (g,\text{RSD}), (g, f_\text{NL})$ for the galaxy clustering, magnification bias, redshift space distortion and primordial local non-Gaussianity respectively, the weak lensing and intrisic alignment contributions are written $i,j = (g, \gamma), \text{IA}$ and we use $i,j = (\text{CMB}, \kappa), (\text{CMB},T)$ for CMB lensing and CMB ISW effect. In general, the source functions are of the form

\begin{equation}
    \Delta_\ell^i(k) = c(\ell)\int_0^\infty \text{d} \chi \ W^i(\chi) j^{(n)}_\ell(k\chi),
\end{equation}
with $c(\ell)$ some known $\ell$-dependent coefficients, $j^{(n)}_\ell(k\chi)$ the $n$-th derivative of the spherical Bessel function and $W^i(\chi)$ the window functions that we define for the cosmological probes implemented in $\texttt{SwiftC}_\ell$ in the following. We choose to split the power spectrum as follows:
\begin{equation}
    P(k,\chi,\chi') = P(k, z=z_\text{fid})D(k,z(\chi))D(k,z(\chi')),
    \label{eq:power spectrum split}
\end{equation}
with $D(k,z)$ defined as 
\begin{equation}
    D(k, z(\chi)) = \sqrt{P(k,z)/P(k,z_\text{fid})},
    \label{eq:D_k}
\end{equation}
where $P(k,z) \equiv P(k, \chi(z), \chi(z))$. This split allows for a mild $k$-dependent $D(k,z)$ and is thus well-suited for the numerical evaluation that we describe in the following. In the case of a linear power spectrum, $D(k, z)$ reduces to the usual definition of the growth factor. The pipeline however also supports models with highly $k$-dependent growth factor. In that case, this splitting allows for most of the $k$-dependence to be carried by $P(k, z = z_\text{fid})$. $z_{\text{fid}}$ can be chosen as the mean redshift within a given redshift distribution such that the residual scale dependence carried by $D(k,z)$ is as mild as possible. 

In the next sections, we describe the window and source functions for the probes implemented in $\texttt{SwiftC}_\ell$.

\subsection{Galaxy clustering}
The window function for galaxy clustering can be expressed as follows:
\begin{equation}
    W^{\delta, g}(\chi)= n(z(\chi)),
\end{equation}
with $n(z(\chi))$ the galaxy redshift distribution normalised to $\int \text{d}z \ n(z) = 1$.
In this work, we assume a linear galaxy bias model as done in \cite{Kaiser1984} and \cite{Bardeen1986}. The contribution to the angular power spectrum is then given by
\begin{equation}
    \Delta_\ell^{g, D}(k) = b_1\int_0^\infty \text{d}\chi \ W^{\delta, g}(\chi)D(k,z(\chi))j_\ell(k\chi),
\end{equation}
with $b_1$ the galaxy bias. For simplicity, we assume a time- and scale-independent effective bias per window function. Note that a more complex non-linear model of bias could seamlessly be included in the pipeline. Other physical contributions can be added to the galaxy clustering kernel, such as the magnification bias, redshift space distortion (RSD) and primordial local non-Gaussianity that we define below. The kernel can then be written as
\begin{equation}
    \Delta_\ell^g = \Delta_\ell^{g, D} + \Delta_\ell^{g,\mu} + \Delta_\ell^{g,\text{RSD}} + \Delta_\ell^{g, f_{\text{NL}}},
\end{equation}
where we define $\Delta_\ell^{g,\mu}$, $\Delta_\ell^{g,\text{RSD}}$ and $\Delta_\ell^{g, f_{\text{NL}}}$ in equations \eqref{eq:magnification bias}, \eqref{eq:RSD} and \eqref{eq:f_NL} respectively.

\subsubsection{Magnification bias}

The tomographic lens efficiency is given by \cite{darkenergysurveyyear3}
\begin{equation}
    W^{\kappa, \alpha}(\chi) = \frac{3\Omega_mH_0^2}{2c^2}\int_\chi^\infty \text{d}\chi' \ n^\alpha(\chi')\frac{\chi}{a(\chi)}\frac{\chi'-\chi}{\chi'},
\label{eq:tomographic lense efficiency}
\end{equation}
with $\alpha = g,\gamma$ for galaxy or lens redshift distribution, $a(\chi)$ the scale factor, $\Omega_m$ the matter density parameter, $H_0$ the Hubble constant at present time and $c$ the speed of light.
The magnification bias contribution at linear order is then defined as
\begin{equation}
    \Delta_\ell^{g,\mu}(k) = \frac{\ell (\ell+1)}{k^2}C_g\int_0^\infty \frac{\text{d}\chi}{\chi^2}\ W^{\kappa,g}(\chi)D(k,z(\chi))j_\ell(k\chi),
\label{eq:magnification bias}
\end{equation}
where $C_g$ is the lensing bias coefficient.

\subsubsection{RSD contribution}
The redshift space distortion (RSD) contribution is given by \cite{darkenergysurveyyear3}
\begin{equation}
    \Delta_\ell^{g,\text{RSD}}(k) = -\int_0^\infty \text{d}\chi \ f(z(\chi))W^{\delta,g}(\chi)D(k,z(\chi))j_\ell^{(2)}(k\chi),
\label{eq:RSD}
\end{equation}
with $f(z) = d\ln D/d\ln a$ the logarithmic growth rate. We assume $f(z)$ to be scale independent but conserve the scale dependence of $D(k, z)$. Note that equation \eqref{eq:RSD} is strictly valid in describing RSD only at linear level. However, since RSDs are vanishing at high $\ell$, plugging a non-linear matter power spectrum into equation \eqref{eq:RSD} introduces tolerably small errors.  

\subsubsection{Primordial local non-Gaussianity}
Assuming a local primordial non-Gaussianity, we can write the galaxy density as follows \cite{Barreira_2022}:
\begin{equation}
    \delta^{g, f_\text{NL}}(k,z) = \frac{b_1^{f_{\text{NL}}}f_\text{NL}}{T_\alpha}\delta_m(k,z) + \epsilon(k,z),
\end{equation}
with $\epsilon(k,z)$ a stochastic or noise variable, $b_1^{f_{\text{NL}}}$ the so-called non-local bias and the transfer function $T_\alpha$ given by 
\begin{equation}
    T_\alpha(k, \chi) = \frac{2}{3}T_m(k)\frac{k^2c^2}{H_0^2\Omega_m}\frac{D(\chi(z))}{D(\chi(z_{\text{MD}}))(1+z_{\text{{MD}}})},
\end{equation}
where $T_m(k)$ is the matter transfer function at redshift $z=0$, $z_{\text{MD}}$ is the redshift at matter domination and $D(z)$ is the usual scale-independent growth factor. 
The contribution of the local primordial non-Gaussianity is then defined as \cite{pierrefnl}
\begin{equation}
    \Delta_\ell^{g, f_{\text{NL}}} = b_1^{f_{\text{NL}}}f_{\text{NL}}\int_0^\infty \text{d}\chi \ \frac{W^{\delta, g}(\chi)}{T_\alpha(k, \chi)}D(k, z(\chi)) j_\ell(k\chi).
\label{eq:f_NL}
\end{equation}

\subsection{Weak gravitational lensing}
The contribution for gravitational lensing is given by \cite{N5KChallenge}
\begin{equation}
    \Delta_\ell^{\gamma}(k) = \sqrt{\frac{(\ell + 2)!}{(\ell-2)!}}\frac{1}{k^2}\int_0^\infty \frac{\text{d}\chi}{\chi^2} \ W^{\kappa, \gamma}(\chi)D(k,z(\chi))j_\ell(k\chi),
\end{equation} with $W^{\kappa, \gamma}$ the window function defined in equation \eqref{eq:tomographic lense efficiency}.
We also include the non-linear alignment model (NLA) whose contribution can be expressed as \cite{alex12x2, pyccl}
\begin{equation}
    \Delta_\ell^\text{IA}(k) = \sqrt{\frac{(\ell+2)!}{(\ell-2)!}}\frac{1}{k^2}\int_0^\infty \frac{\text{d}\chi}{\chi^2} \ n^\gamma(\chi)A_\text{IA}(z(\chi))D(k,z(\chi))j_\ell(k\chi),
\end{equation}
with $A_\text{IA}(z(\chi))$ the NLA amplitude parameter.

\subsection{CMB lensing}
The contribution for CMB lensing is very similar to the weak gravitational lensing with the difference that the integral over $\chi$ is evaluated up to the last scattering surface at $\chi_*$. It can be written as \cite{Durrer_2020}
\begin{equation}
    \Delta_\ell^{\text{CMB}, \kappa}(k) = \ell(\ell+1)\frac{3\Omega_mH_0^2}{2c^2k^2}\int_0^{\chi_*} \frac{\text{d}\chi}{a(\chi)}\frac{\chi_*-\chi}{\chi\chi_*}D(k,z(\chi))j_\ell(k\chi).
\end{equation}

\subsection{CMB Integrated Sachs-Wolfe (ISW) effect}
The window function for the Integrated Sachs-Wolfe effect can be expressed as follows \cite{Nicola_2016}
\begin{equation}
    W^{\text{CMB},T} = 2T_\text{CMB}\int_{\eta_*}^0 d\eta \frac{d\Phi}{d\eta},
\end{equation}
with $T_\text{CMB}$ the temperature of the CMB at present time, $\Phi$ the gravitational potential and $\eta_*$ the conformal time of the last scattering surface. The kernel can then be written as
\begin{equation}
    \Delta_\ell^{\text{CMB},T}(k) = 2T_\text{CMB} \frac{3\Omega_mH_0^2}{2c^3k^2}\int_0^{\chi_*}\text{d}\chi \ (1-f(z(\chi)))D(z(\chi))H(z(\chi))j_\ell(k\chi).
\end{equation} 
Note that for simplicity, we assume a fully scale-independent growth rate for this probe as this is sufficiently accurate for this large-scale signal.
 
\subsection{Limber approximation}
In the Limber approximation \cite{Limber}, the spherical Bessel functions are assumed to be Dirac delta functions, i.e.\ $j_\ell(k \chi) \simeq \sqrt{\frac{\pi}{2\ell + 1}} \delta^D(\ell + \frac{1}{2} - k \chi)$. The angular power spectrum then reduces to a single integral over $\chi$:

\begin{equation}
    C_\ell^{ij} \approx \int_0^\infty \frac{\text{d}\chi}{\chi^2} \ W^i(\chi) W^j(\chi) P\bigg(k = \frac{\ell + \frac{1}{2}}{\chi}, \chi, \chi\bigg).
\label{eq:Limber}
\end{equation}
The error of this approximation is of order $\mathcal{O}(\ell^{-1})$ and is thus only useful on small angular scales.

\section{Computation of the angular power spectrum}
\label{sec:angular power spectrum}
\subsection{FFTLog}
The FFTLog method relies on the Fast Fourier Transform (FFT) algorithm in logarithmic space. For a function $f(\chi)$, it is defined as \cite{Hamilton_2000}

\begin{equation}
    f(\chi) \approx \sum_{n = -N_\text{FFT}/2}^{N_\text{FFT}/2} c_n \ \chi^{b + i\eta_n},
\label{eq:FFTLog}
\end{equation}
with $N_{FFT}$ the number of points and $\eta_n = \frac{2\pi n}{\ln(\chi_\text{max}/\chi_\text{min})}$.
The coefficients $c_n$ can be expressed as
\begin{equation}
    c_n = \frac{1}{N_\text{FFT}}\sum_{m = 0}^{N_\text{FFT}+1}f(\chi_m)\chi_m^{-b}\chi_\text{min}^{-i\eta_n}e^{-2inm/N_\text{FFT}}.
\end{equation}
The FFTLog method can be subject to numerical instabilities such as aliasing and ringing. To minimise these effects, we introduce a smoothing function at the tails and introduce a bias $b$. The bias $b$ can be set to any non-integer real number but is usually chosen with small magnitude for a better convergence.

\subsection{Computational Method}
In this work, we opt for an FFTLog decomposition of the $\chi$-dependent part of the computation, similar to what is done in \cite{Fang_2020}, but we decide to keep the $k$-dependence by computing $k$-dependent Fourier coefficients.
As shown in equations \eqref{eq:power spectrum split} and \eqref{eq:D_k}, we first decompose the power spectrum in the following way:
\begin{equation*}
    P(k,\chi,\chi') = P(k, z=z_\text{fid})D(k,z(\chi))D(k,z(\chi')),
\end{equation*}
with $D(k, z(\chi)) = \sqrt{P(k,z)/P(k,z_\text{fid})}$. We then take the FFTLog decomposition of the $\chi$-dependent functions: 
\begin{equation}
    W(\chi)D(\chi, k) = \sum_{n = - N_\text{FFT}/2}^{N_\text{FFT}/2}c_{n}(k)\ \chi^{b + i\eta_{n}},
\end{equation}
for different points in $k$-space, where $b$ and $\eta_n$ are defined as in equation \eqref{eq:FFTLog}.
The integral over $\chi$ (or over $\chi'$) then has an analytical solution \cite{Table_of_Integrals}: 
\begin{equation}
    \int_0^\infty \text{d}\chi \ \chi^p j_\ell(k\chi) = 2^{-1+p} k^{-1-p}\sqrt{\pi}\frac{\Gamma{(\frac{1}{2}(1+\ell+p}))}{\Gamma({\frac{1}{2}(2+\ell+p)})},
\label{eq:int_jl}
\end{equation}
where we assume $k>0$ and $-\Re(\ell)-1 < \Re(p)<1$. The gamma ratios depend only on $\ell$ and $p$ and can thus be precomputed for fixed $\ell$'s and a fixed number of FFTLog points.
We finally perform a numerical integral over $k$:

\begin{equation}
    C_\ell \approx \sum_{p,q} 2^{-1+p+q}\frac{\Gamma{(\frac{1}{2}(1+\ell+p}))}{\Gamma({\frac{1}{2}(2+\ell+p)})} \frac{\Gamma{(\frac{1}{2}(1+\ell+q}))}{\Gamma({\frac{1}{2}(2+\ell+q)})} \int_0^\infty \text{d}k \ c_p(k) c_q(k) k^{-p-q} P(k).
\end{equation}
For a faster computation, we choose to compute the Fourier coefficients for \texttt{N$_\texttt{interp}$} points in $k$-space and interpolate the rest of the coefficients. This is possible thanks to the split in the power spectrum and the mild $k$-dependence of $D(k,z)$. 

The numerical hyper-parameters of $\texttt{SwiftC}_\ell$ are thus \texttt{N$_\texttt{FFT}$}, the number of points in the FFTLog, and \texttt{N$_\texttt{interp}$}, the number of points in the interpolation. In addition to these internal parameters, the integral over $k$ is automatically performed over all \texttt{N$_\texttt{k}$} points on which the input power spectrum provided by the user is defined. The impact of these three parameters on the performance is explored in further detail in the Appendix \ref{appendix}.

In recent years, various beyond-Limber methods have been developed. Several methods also employ the FFTLog algorithm, as can be seen for example in \texttt{FKEM} and \texttt{matter} from the N5K challenge \cite{N5KChallenge}. An alternative method, \texttt{Levin}, computes the oscillatory integrals as solutions to a system of differential equations that can be solved using linear algebra. Finally, \texttt{AngPow} \cite{Angpow} and \texttt{Blast.jl} \cite{blast} make use of the Chebyshev polynomials, either to optimise quadrature in the former or as a decomposition basis in the latter. A comparison of the main characteristics of different codes can be found in Table \ref{tab:code_comparison}. An important feature to take into account when considering the different algorithms is the incorporation of $k$-dependent growth (Cf. last column of the Table). For example, \texttt{FKEM} opts for a splitting between the linear and non-linear power spectrum, where the non-linear contribution is then evaluated using the Limber approximation, whereas in $\texttt{SwiftC}_\ell$, \texttt{Levin}, \texttt{matter} or \texttt{Blast.jl} this splitting is not necessary and these codes deal with the full power spectrum. \texttt{FKEM} further assumes the growth factor to be $k$-independent for the linear part of the power spectrum. This factorisation does not allow for the analysis of probes with relevant $k$-dependence at large scales, such as primordial non-Gaussianities. Furthermore, this split becomes increasingly inaccurate in beyond-$\Lambda$CDM models such as models with massive neutrinos or modified gravity.

\section{Validation of implementation}
\label{sec:Results}

\subsection{Overview of all the available probes}
\label{sec:CCL}
\begin{figure}[ht]
\centering
\includegraphics[width=\textwidth]{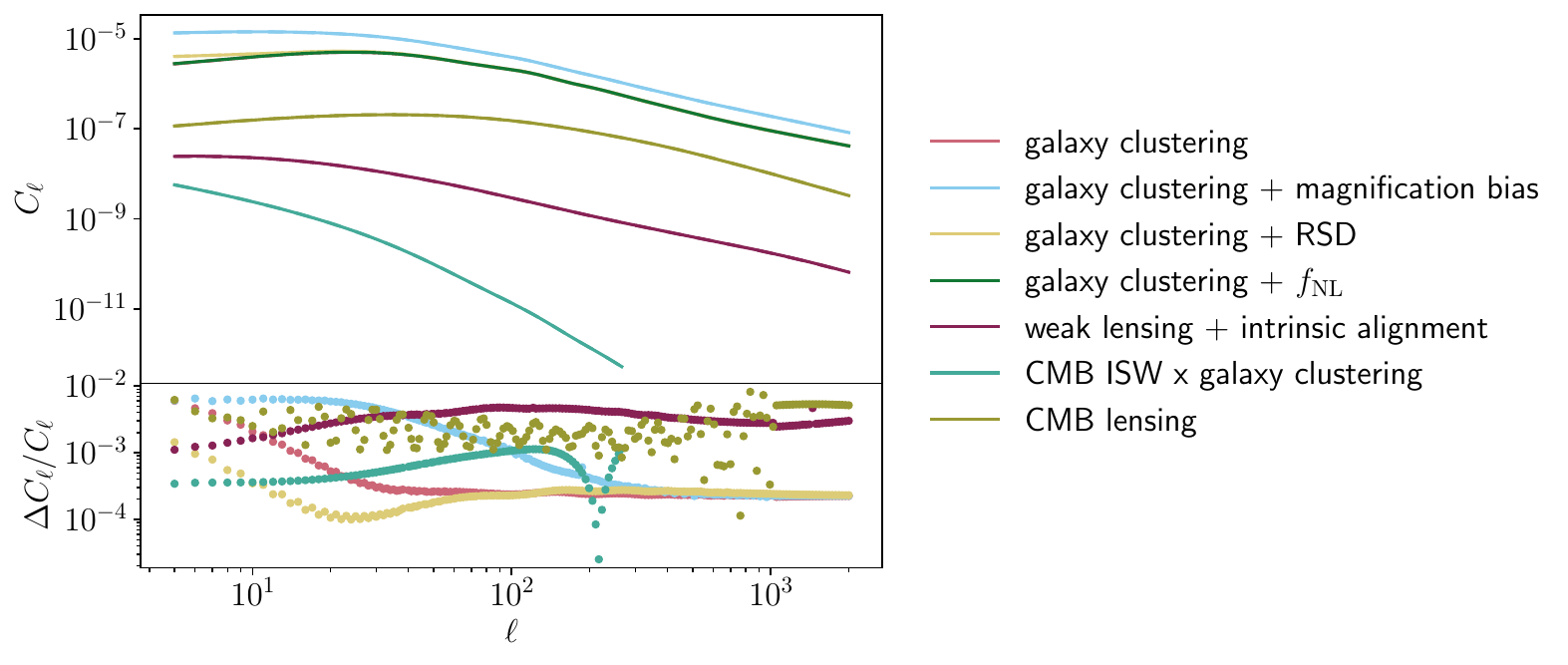}
\caption{All probes available in $\texttt{SwiftC}_\ell$ compared to the package \texttt{CCL} when possible. In \texttt{CCL}, we use \texttt{FKEM} for $\ell < 1000$ and then switch to the Limber approximation. We find that the pipelines agree to subpercent level.}
\label{fig:All_contr}
\end{figure}

We plot the available probes, i.e.\ galaxy clustering, including magnification bias, redshift-space distortions and primordial non-Gaussianity, weak lensing with intrinsic alignment, CMB ISW $\times$ galaxy clustering and CMB lensing against the \texttt{CCL} package \cite{pyccl} in Figure \ref{fig:All_contr}. In \texttt{CCL}, we use \texttt{FKEM} for $\ell < 1000$ and then switch to the Limber approximation for higher $\ell$'s. The values of the different parameters can be found in Table \ref{tab:fiducial_cosmology_CCL}. We find that $\texttt{SwiftC}_\ell$ agrees with \texttt{CCL} to subpercent level for all probes.

\subsection{N5K Challenge}
\label{sec:N5K}
\subsubsection{Set-up}
The N5K challenge's\footnote{\url{https://github.com/LSSTDESC/N5K.git}} goal was to encourage scientists to develop faster and more accurate tools for the computation of the angular power spectrum in anticipation of LSST data. The challenge consists in computing $3\times2$pt angular power spectra, i.e.\ galaxy clustering and weak lensing auto- and cross-correlations, for 10 galaxy clustering kernels and 5 weak lensing kernels, amounting to a total of 120 different angular power spectra for 103 log-spaced multipoles $\ell$ from 2 to 2000. The team behind the challenge computed benchmarks using brute-force integration in order to quantify the accuracy of the different entries. Initially, the accuracy requirement was of $\Delta \chi^2 < 1$ for the whole $\ell$-range however, due to all submitted pipelines switching to Limber after $\ell = 200$, the accuracy requirement was adjusted to its equivalent of $\Delta \chi^2 < 0.2$ for $\ell < 200$.

\begin{table}[]
    \centering
    \begin{tabular}{ l | c | c | p{9.0em} | C{4.0em} | p{6.5em} }
         \textbf{Software} & \textbf{Language} & \textbf{Auto-diff} & \textbf{Probes} & \textbf{Beyond Limber} & \textbf{$k$-dependent growth}  \\
         \hline \hline
         $\texttt{SwiftC}_\ell$ & \texttt{JAX} & Yes & Galaxy clustering with magnification bias, RSD, PNG; weak lensing with IA, CMB lensing, CMB ISW & Yes & Exact\\[5ex] \hline 
         \texttt{Blast.jl} & \texttt{Julia} & Yes & Galaxy clustering, weak lensing & Yes & Exact\\ \hline
         \texttt{JAX-Cosmo} & \texttt{JAX} & Yes & Galaxy clustering, weak lensing with IA & No & No \\ \hline
         \texttt{LimberJack.jl} & \texttt{Julia} & Yes & Galaxy clustering, weak lensing with IA, CMB lensing & No & No \\ \hline
         \texttt{CCL} (\texttt{FKEM}) & \texttt{Python} & No & Galaxy clustering with magnification bias, RSD; weak lensing with IA, CMB lensing, CMB ISW, CIB, tSZ\footnotemark & Yes & Approximate
    \end{tabular}
    \caption{A table of different available angular power spectra codes.}
    \label{tab:code_comparison}
\end{table}
\footnotetext{CIB refers to the cosmic infrared background and tSZ to the thermal Sunyaev Zel’dovich effect.}

\begin{table}[ht]
\centering
\begin{tabular}{lS[table-format = 2.3]}
\hline
\textbf{Parameter} & \textbf{Fiducial value} \\ \hline
$\omega_b$                        & 0.022 \\
$\omega_{\mathrm{cdm}}$           & 0.120 \\
$h$                                   & 0.678 \\
$n_s$                                 & 0.966 \\
$\ln(10^{10} A_s)$                    & 3.045 \\
$b_1$                   & 1.000 \\ 
$C_g$ & 18.000 \\
\hline
\end{tabular}
\caption{Fiducial cosmological parameter set used in the test presented in section $\S$\ref{sec:CCL}.}
\label{tab:fiducial_cosmology_CCL}
\end{table}

Three different entries were submitted for this challenge, \texttt{FKEM}, \texttt{Levin} and \texttt{matter}. In this context, $\texttt{SwiftC}_\ell$ stands out thanks to the exact treatment of the $k$-dependent growth factor and its efficiency. Indeed, $\texttt{SwiftC}_\ell$ can account for additional scale dependence that can be found for example in beyond-$\Lambda$CDM models such as models with massive neutrinos or modified gravity. Importantly, it does not require an approximate splitting in the power spectrum between an approximately factorisable linear term and a residual non-linear correction piece, where e.g. the latter is computed using the Limber approximation in \texttt{FKEM}. $\texttt{SwiftC}_\ell$ simply deals with the full power spectrum thanks to a smooth interpolation in $k$ between line-of-sight integrals computed from $k$-dependent FFTLog decompositions. 

\subsubsection{Results}
\begin{figure}[ht]
\centering
\includegraphics[width=\textwidth]{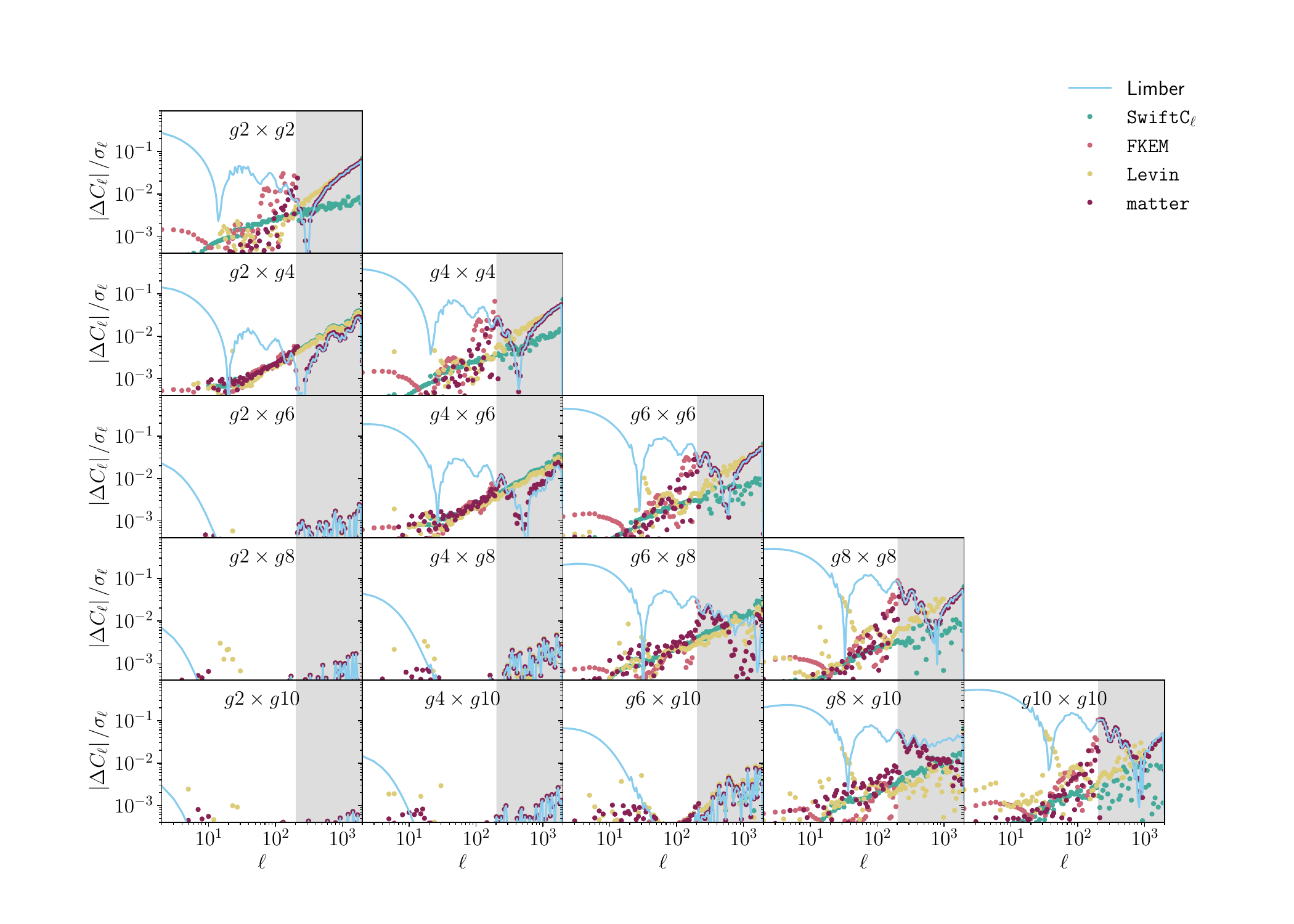}
\caption{Accuracy of the galaxy clustering auto- and cross-correlated angular power spectra for half of the bins of the N5K challenge for all the entries and $\texttt{SwiftC}_\ell$ against the benchmarks. The uncertainties $\sigma_\ell$ represent a simple Gaussian covariance and the grey region represents the $\ell>200$ range that is not part of the challenge.}
\label{fig:N5K_gg}
\end{figure}

\begin{figure}[ht]
\centering
\includegraphics[width=\textwidth]{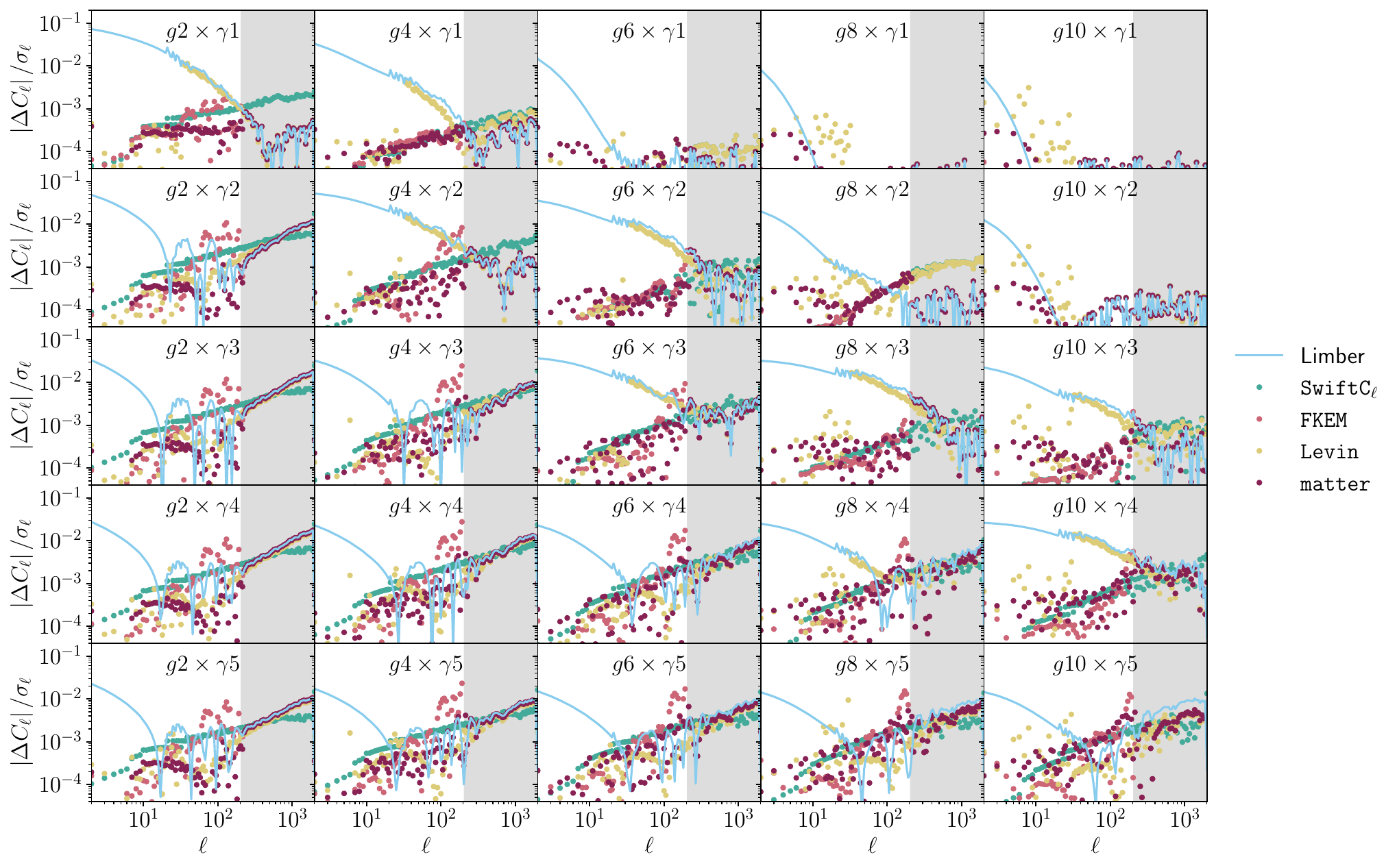}
\caption{Relative difference in the galaxy-galaxy lensing cross-correlated angular power spectra for some of the bins of the N5K challenge for all the entries and $\texttt{SwiftC}_\ell$ against the benchmarks. The uncertainties $\sigma_\ell$ represent a simple Gaussian covariance and the grey region represents the $\ell>200$ range that is not part of the challenge.}
\label{fig:N5K_gls}
\end{figure}

\begin{figure}[ht]
\centering
\includegraphics[width=\textwidth]{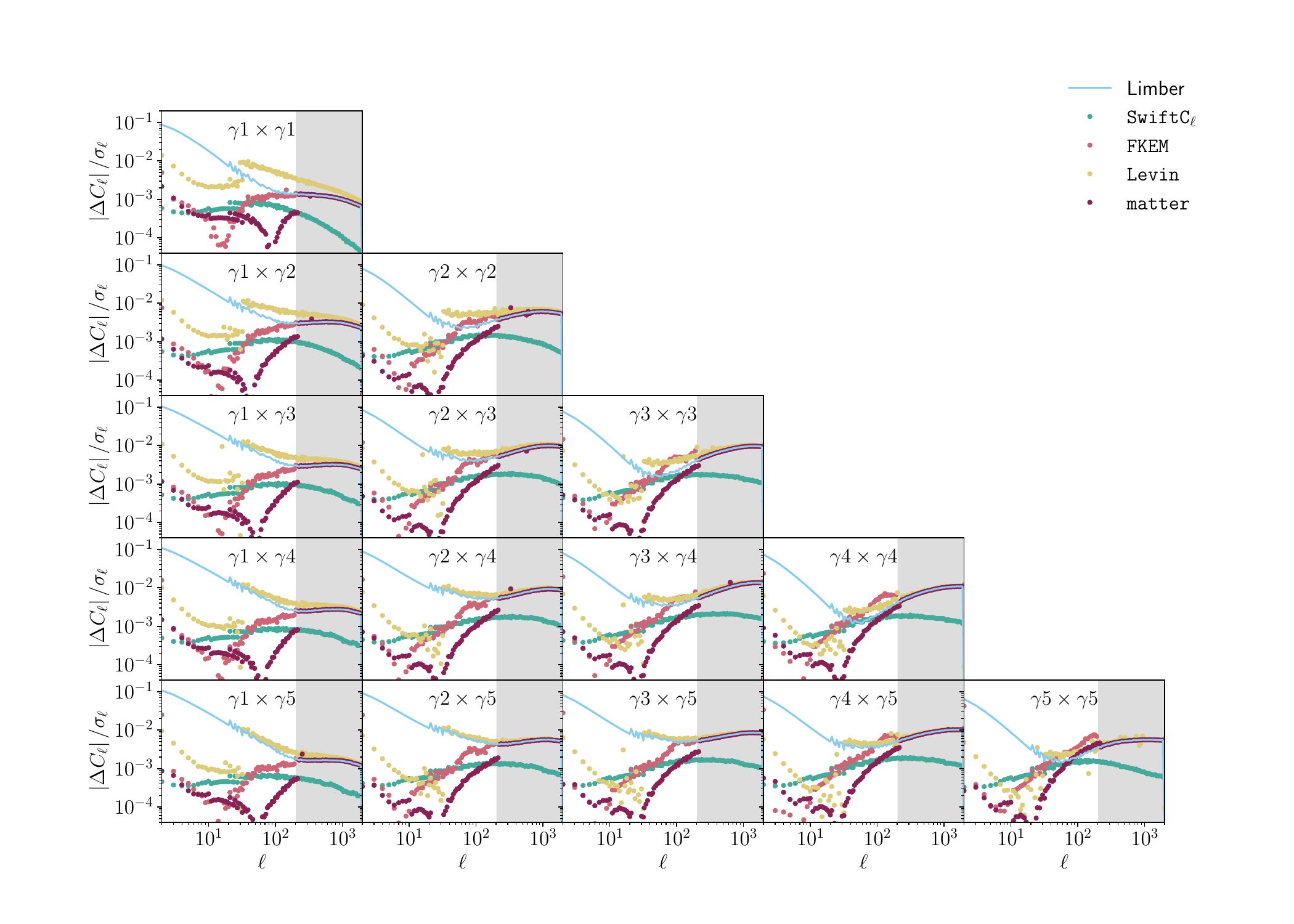}
\caption{Relative difference of the weak lensing auto- and cross-correlated angular power spectra for all of the bins of the N5K challenge for all the entries and $\texttt{SwiftC}_\ell$ against the benchmarks. The uncertainties $\sigma_\ell$ represent a simple Gaussian covariance and the grey region represents the $\ell>200$ range that is not part of the challenge.}
\label{fig:N5K_ls}
\end{figure}

We run the $3\times2$pt analysis for $\texttt{SwiftC}_\ell$ and the previous challenge entrants on both a single GPU core on a Nvidia Tesla A100 node and a single CPU core on an AMD EPYC 7742 node\footnote{\url{https://scicomp.ethz.ch/wiki/Getting_started_with_GPUs}}. It is important to note that contrary to the entries in the N5K challenge, we have not implemented a switch to Limber approximation and use our FFTLog method for the whole $\ell$ range. For this reason, we choose our parameters so that $\texttt{SwiftC}_\ell$ fulfils the requirement of $\chi^2 < 1$ over the whole $\ell$ range, whereas the other pipelines of the challenge use settings that meet an equivalent accuracy requirement of $\chi^2 < 0.2$ for $\ell < 200$. The results for all the pipelines can be found in Table \ref{tab:N5K}. When running on a GPU core, we find that $\texttt{SwiftC}_\ell$ performs the whole analysis in $0.006$ s, i.e.\ approximately 40$\times$ faster than the winner of the challenge, \texttt{FKEM}. On CPU, the code is slower, as expected and runs in 0.5 s, about twice the timing of \texttt{FKEM}. The running times do not take into account the pre-computation of the gamma function ratios from equation \eqref{eq:int_jl} and the compiling of the pipeline as these are only run once and as allowed from the challenge. The accuracy of the different angular power spectra for some of the redshift bins of the different pipelines and the Limber approximation can be found in Figures \ref{fig:N5K_gg}, \ref{fig:N5K_gls} and \ref{fig:N5K_ls} for galaxy clustering, galaxy-galaxy lensing and weak lensing respectively. We can see that $\texttt{SwiftC}_\ell$ satisfies effortlessly the accuracy requirements. Furthermore, it is more accurate than the Limber approximation even in the high $\ell$-range for all the galaxy clustering and weak lensing power spectra shown.

\begin{table}
\centering
\begin{tabular}{c|c|S[table-format = 2.3(1), uncertainty-mode = separate, retain-zero-uncertainty = true]}
    Node & Entry name & {Runtime [s]} \\
    \hline
    GPU & $\texttt{SwiftC}_\ell$ & 0.006(0.000) \\
    \hline
    \multirow{4}{*}{CPU} & $\texttt{SwiftC}_\ell$ & 0.573(0.009) \\
    & \texttt{FKEM} & 0.249(0.029)  \\
    & \texttt{matter} & 3.136(0.018) \\
    & \texttt{Levin} & 10.940(0.101)  \\
\end{tabular}
\caption{Running times for the N5K Challenge entries and $\texttt{SwiftC}_\ell$ using a single GPU core and a single CPU core respectively. The parameters of all the different pipelines are adjusted to fulfil $\Delta \chi^2 < 0.2$ for $\ell < 200$, except for $\texttt{SwiftC}_\ell$ for which we fulfil $\Delta \chi^2 < 1$ over the whole $\ell$-range. As can be seen, $\texttt{SwiftC}_\ell$ greatly benefits from running on a GPU node rather than on a CPU node. On a GPU node, $\texttt{SwiftC}_\ell$ is approximately 40$\times$ faster than the winner of the challenge, \texttt{FKEM} for comparable accuracy.}
\label{tab:N5K}
\end{table}

The N5K Challenge also provides half- and quarter-width bins. The set-up used for the different bin widths can be found in Table \ref{tab:N5K_bins}. We find that the FFTLog decomposition requires more points to properly reconstruct thinner window functions in order to meet the accuracy requirements, as expected. The running times for these set-ups are $\sim40$ ms and $\sim10$ ms for the half- and quarter-width bin respectively.

\begin{table}[h]
    \centering
    \begin{tabular}{c|ccc|c}
        Bin width & \texttt{N$_\texttt{FFT}$} & \texttt{N$_\texttt{interp}$} & \texttt{N$_{k}$} & $\chi^2$ \\
        \hline
        Full & 512 & 200 & 300 & 0.550   \\
        Half & 2048 & 150 & 1000 & 0.814   \\
        Quarter & 1024 & 200 & 500 & 0.923   \\ 
    \end{tabular}
    \caption{Set-up for different bin widths. $\texttt{SwiftC}_\ell$ reaches the desired accuracy effortlessly by tuning in \texttt{N$_\texttt{FFT}$}, \texttt{N$_\texttt{interp}$} and \texttt{N$_{k}$}.}
    \label{tab:N5K_bins}
\end{table}

\subsection{MCMC, HMC and Fisher information matrix}
\label{sec:MCMC}
\subsubsection{Set-up}

\begin{table}[ht]
\centering
\begin{tabular}{lS[table-format = 2.3]}
\hline
\textbf{Parameter} & \textbf{Fiducial value} \\ \hline
$\omega_b$                        & 0.022 \\
$\omega_{\mathrm{cdm}}$           & 0.120 \\
$h$                                   & 0.672 \\
$n_s$                                 & 0.964 \\
$\ln(10^{10} A_s)$                    & 3.060 \\
$b_1$                   & 1.000 \\ \hline
\end{tabular}
\caption{Fiducial cosmological parameter set used in the analysis presented in section $\S$\ref{sec:MCMC}.}
\label{tab:fiducial_cosmology}
\end{table}

To demonstrate the utility of $\texttt{SwiftC}_\ell$, we run an ensemble sampling as implemented in \texttt{emcee} \cite{emcee}, a Hamiltonian Monte Carlo and a Fisher forecast using a simple forecast mock data vector and covariance from LSST following the LSST Science Requirements Document (SRD) \cite{LSST_SRD}. The mock data vector is computed using the \texttt{CCL} package \cite{pyccl} and LSST forecast Y10 redshift bins for galaxy clustering and weak lensing using a simple linear bias model for galaxy clustering and the usual weak lensing kernel without intrinsic alignment. The set-up includes 10 auto-correlation bins for galaxy clustering, 15 tomographic redshift bin combinations for the 5 weak lensing redshift bins and 25 galaxy-galaxy lensing cross-correlations. The lens-source bin combinations are only included in the data vector if the lens bin is at lower redshift than the sources, allowing for at most a 10\% overlap. We use an $\ell$-range of $2<\ell<2000$ and use 20 log-spaced bins for each spectrum in order to test the low-$\ell$ region where beyond-Limber corrections become important. We note that this does not represent a realistic range of scales for the LSST analysis but is sufficient to demonstrate the accuracy of $\texttt{SwiftC}_\ell$ in an analysis setting. We compute a covariance matrix using a Gaussian approximation for the cosmic variance term and additive noise terms computed using the forecast number density of the galaxy samples and ellipticity dispersion (see \cite{LSST_SRD} for details of the forecasted number density and ellipticity dispersion). We use a simple $1/f_\text{sky}$ approximation in computing the covariance, data vector and theoretical predictions to account for the effect of masking with $f_\text{sky} = 0.4363$ the fraction of the sky to be covered by LSST. We use the fiducial cosmology shown in Table \ref{tab:fiducial_cosmology} and fix all parameters to their fiducial values aside from $\omega_{\mathrm{cdm}}$, $\ln(10^{10} A_s)$ and $h$  which we vary in the analysis over a wide, flat prior. For simplicity, we fix the nuisance parameters such as linear biases and ignore magnification bias, RSD and intrinsic alignment, making our settings unrealistic but sufficient for testing the pipeline. We make use of \texttt{CosmoPower-JAX} \cite{CPJ} to compute the non-linear matter power spectrum and write our likelihood in pure \texttt{JAX}, using \texttt{vmap} to vectorise the likelihood. We use \texttt{BlackJAX} \cite{BlackJAX} for the HMC. The contour plots are realised using \texttt{GetDist} \cite{getdist}. 

\subsubsection{Results}
\begin{figure}[ht]
\centering
\includegraphics[width=0.7\textwidth]{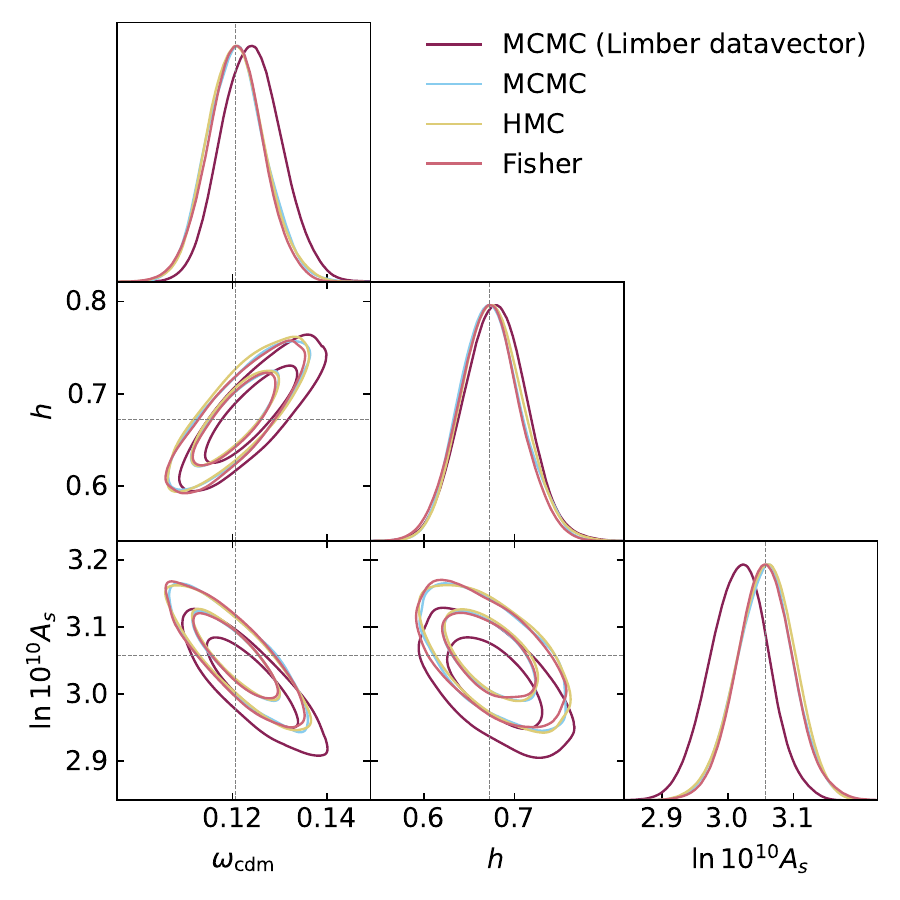}
\caption{Ensemble sampling as implemented in \texttt{emcee}, HMC and a Fisher forecast with data vector computed using \texttt{CCL} with \texttt{FKEM} unless stated otherwise. As can be seen, the parameters are fully recovered in the ensemble sampling when using \texttt{FKEM}, in the HMC and in the Fisher forecast whereas a slight shift in some parameters can be observed in the ensemble sampling when using the Limber approximation. The Fisher information matrix is computed using the auto-differentiation from the \texttt{JAX} package.}
\label{fig:MCMC}
\end{figure}

We compute the data vector using \texttt{CCL} with \texttt{FKEM} for ensemble sampling, HMC and Fisher matrix. We run a second ensemble sampling, with a data vector with the Limber approximation using \texttt{CCL}. We also make use of the auto-differentiation from \texttt{JAX} to compute the Fisher information matrix as the Hessian of the log-likelihood. As can be seen in Figure \ref{fig:MCMC}, the ensemble sampling with the \texttt{FKEM} data vector, the HMC and the Fisher information recover the parameters fully whereas the ensemble sampling with the data vector computed via the Limber approximation has some shift in the parameters. We perform the analysis on GPU Nvidia Tesla A100 node using a single core. The ensemble sampling is ran with 12 walkers with 2000 points each in 162s whereas the HMC is ran using the no-U-turn (NUTS) algorithm with 2000 points on a single chain in 23min34s.

\subsubsection{Discussion}
We benchmark the likelihood and its gradient in this simplified analysis set-up. A call of the likelihood takes $\sim 6.74$ms whereas its gradient takes $\sim 28.7$ms. The gradient is thus approximately $4\times$ slower than the likelihood itself. If we assume the use of the five-point stencil method for the computation of finite differences, the auto-differentiation running time will thus be of the same order (or slightly faster). Furthermore, auto-differentiation does not suffer from numerical instability contrary to finite differences. 

We now wish to compare ensemble sampling and HMC methods. The evaluation of one point for HMC requires the computation of the likelihood and its gradient and is thus approximately $5\times$ slower than the evaluation of one point for ensemble sampling. This is usually countered by a higher acceptance rate for HMC. In this setting, we find an acceptance rate of 97\% for HMC and 64\% for ensemble sampling. It is worth noting that the acceptance rate of ensemble sampling is usually lower but the simplistic set-up of the analysis likely simplifies the exploration of the parameter-space. We then compute the effective sample size (ESS) and the effective sample rate. The effective sample size is defined as:

\begin{equation}
    \text{ESS} = \frac{N_\text{samples}}{2\tau},
\end{equation}
with $N_\text{samples}$ the number of samples and $\tau$ the integrated autocorrelation time. We compute the ESS using \texttt{GetDist} \cite{getdist}. The effective sample rate is then computed by dividing by the wall time of the computation. In this case, we do not take into account the warm-up phase nor the JIT compilation for both methods. For ensemble sampling, we obtain $\text{ESS/s} \sim 2.37$ whereas we obtain $\text{ESS/s} \sim 0.45$ for HMC. In this simplified setting, we thus find that the gain from fast convergence speed of HMC over ensemble sampling is balanced out by the additional cost in computing the gradient at each step of HMC. HMC will however be more efficient for likelihoods with very high-dimensional or complex parameter-space in which the acceptance rate of ensemble sampling will drop drastically. Similar conclusions were drawn in \cite{pybird} and \cite{Mootoovaloo:2024lpv}.

\section{Conclusion}
\label{sec:Conclusion}

In the era of high-precision cosmology with surveys such as DESI \cite{DESI}, LSST \cite{LSST} or Euclid \cite{Euclid} offering high precision and broad measurements, we introduced $\texttt{SwiftC}_\ell$, a \texttt{JAX}-python differentiable pipeline for the beyond-Limber computation of the angular power spectrum. The pipeline allows for the computation of auto- and cross-correlations of several different probes, such as galaxy clustering, including magnification bias, redshift-space distortions and primordial non-Gaussianity, weak lensing, including intrinsic alignment, CMB lensing and CMB integrated Sachs-Wolfe effect. The use of numerical methods such as the FFTLog decomposition and interpolation allows for arbitrary precision while keeping the full $k$- and $\chi$-dependence beyond the cases where these can be factorised.

The pipeline was compared to the N5K challenge and proved to be around 40$\times$ faster as the best original entry of the challenge at a precision better than a fraction of the expected uncertainties for LSST. In this setting, we ran several tests on the different parameters of $\texttt{SwiftC}_\ell$ in order to quantify the impact of these parameters on the accuracy and the running time. We made use of these tests to choose meaningful values for the default parameters. We also tested the impact of the number of multipoles $\ell$ computed and found that the timing scales as $t \sim (0.003N_\ell + 5)$ ms thanks to vectorisation, in particular reducing overheads. When possible, the probes were also compared with the \texttt{CCL} package, showing agreement at subpercent level. Finally, to display the main usage of the code, we also ran a mock MCMC using \texttt{emcee}, a HMC and a Fisher forecast on LSST-like data and showed that we recover the input parameters, leveraging differentiability and speed brought by $\texttt{SwiftC}_\ell$.

Looking ahead, we believe $\texttt{SwiftC}_\ell$ will prove to be a useful tool for future LSS analyses. For example, $\texttt{SwiftC}_\ell$ allows for rapid Fisher forecasting by making use of the auto-differentiation. Furthermore, this pipeline is ideal for primordial non-Gaussianity analyses that aim to detect its imprint at large angular scales and analysis of the CMB ISW effect where the signal-to-noise ratio peaks at low-$\ell$. These effects are expected to be detected at unprecedented significance in upcoming data from wide-field surveys, making $\texttt{SwiftC}_\ell$ a valuable tool to achieve robust and rapid analyses of the Universe at the large scales. 

\acknowledgments
We thank Shu-Fan Chen for the scientific exchanges. We also thank Andrina Nicola for her guidance and her helpful comments. We use \texttt{numpy} \cite{numpy}, \texttt{scipy} \cite{scipy} and \texttt{matplotlib} \cite{matplotlib} in this work.
PZ acknowledges support from Fondazione Cariplo under the grant No 2023-1205.

\appendix
\section{Numerical accuracy convergence of $\texttt{SwiftC}_\ell$}
\label{appendix}
\begin{figure}[ht]
\centering
\includegraphics[width=0.8\textwidth]{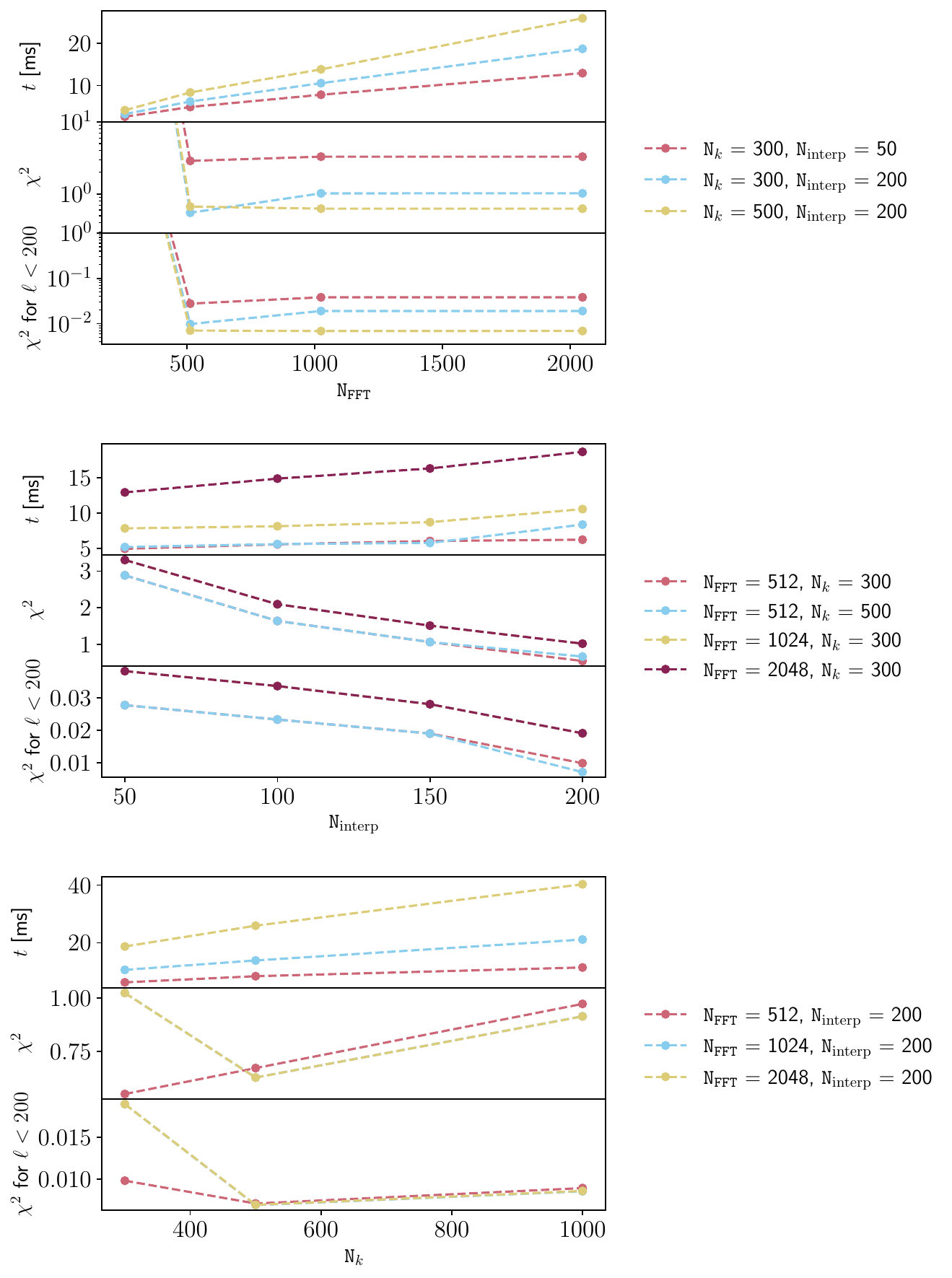}
\caption{Impact of the different parameters \texttt{N$_\texttt{FFT}$}, \texttt{N$_\texttt{interp}$} and \texttt{N$_{k}$} on the running time and accuracy for $\ell$ < 200 and for the whole $\ell$ range in the N5K challenge. Note that in the second and third panels of the first plot, the y-axis is log-scaled.}
\label{fig:timings}
\end{figure}

\begin{figure}[ht]
\centering
\includegraphics[width=0.6\textwidth]{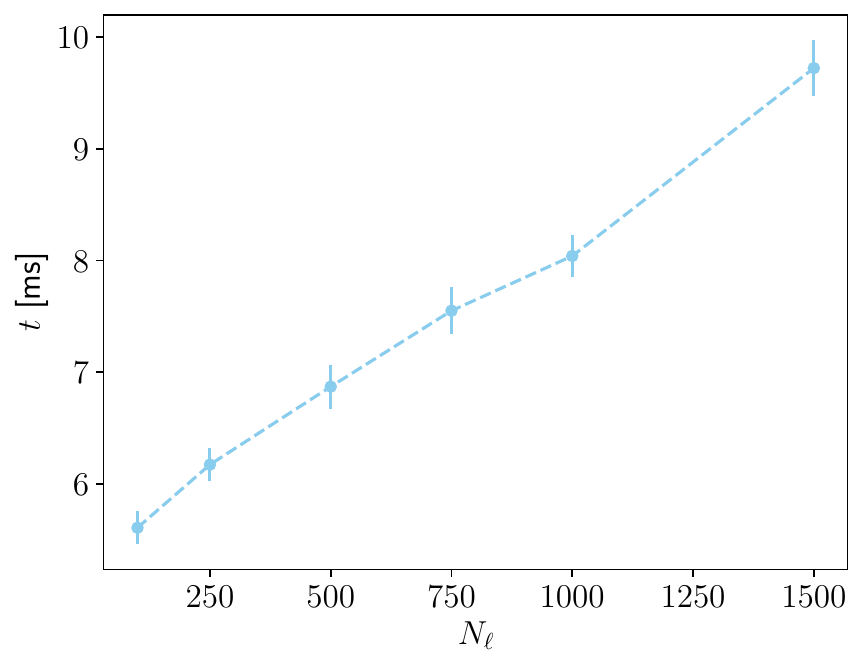}
\caption{Running time as a function of the number of computed $\ell$'s.}
\label{fig:timing_l}
\end{figure}

The performance of $\texttt{SwiftC}_\ell$ can be tailored by changing \texttt{N$_\texttt{FFT}$}, \texttt{N$_\texttt{interp}$} and \texttt{N$_{k}$}. \texttt{N$_\texttt{FFT}$} represents the number of points in the FFTLog decomposition, \texttt{N$_\texttt{interp}$} the number of points in the interpolation along the $k$-range and \texttt{N$_{k}$} the number of points in the $k$-integral. The default configuration is set up with \texttt{N$_\texttt{FFT}$} $= 512$ and \texttt{N$_\texttt{interp}$} $= 200$ for an optimal balance between accuracy and timing in light of tests shown in Figure \ref{fig:timings}. We study the impact of these parameters on the running time and accuracy in the set-up of the N5K challenge \cite{N5KChallenge} from the LSST Collaboration by varying them separately. In this challenge, the speed and the accuracy are tested for the computation of 120 different angular power spectra for galaxy clustering and weak lensing auto- and cross-correlations with 103 $\ell$-multipoles spanning $2 < \ell < 2000$. The team behind the challenge provides benchmark angular power spectra, allowing us to determine the accuracy of $\texttt{SwiftC}_\ell$. The requirements on the accuracy are of $\chi^2<1$ for the whole $\ell$-range and $\chi^2 < 0.2$ for $\ell < 200$. Further details about this challenge are given in section $\S$\ref{sec:N5K}. When increasing the \texttt{N$_\texttt{FFT}$} parameter, we find that the accuracy increases rapidly and then plateaus, showing that the error of the FFTLog decomposition of the window function becomes subdominant. \texttt{N$_\texttt{interp}$} and \texttt{N$_{k}$} mostly impact the accuracy at higher-$\ell$ where the $k$-dependence of the growth factor becomes increasingly important. Furthermore, an interesting interaction between \texttt{N$_\texttt{interp}$} and $\texttt{N}_k$ can be observed: the accuracy decreases when keeping \texttt{N$_\texttt{interp}$} constant while increasing \texttt{N$_{k}$}. This might be due to a decrease in the ratio $\texttt{N}_\texttt{interp}/\texttt{N}_k$, translating into a lower sampling in $k$-space and thus a decrease in the reconstruction accuracy of $k$-dependent growth. The timing of all parameters scales linearly with the tunable parameters. Finally, we test the scaling of the timing as a function of the number of multipoles $N_\ell$ computed, as can be seen in Figure \ref{fig:timing_l}. We find that the timing scales as $t \sim (0.003N_\ell + 5)$ ms. Since the Bessel function is absorbed in the precomputation, $\ell_\text{max}$ has no influence on the effective running time.

As a guideline for the user, the parameters should be tuned in the following way: the parameter that affects the accuracy the most is the number of points in the FFTLog decomposition \texttt{N$_\texttt{FFT}$}. Indeed, a too small \texttt{N$_\texttt{FFT}$} will result in the window function not being accurately reconstructed and propagating error in the computation. For wide bins, the default value is usually sufficient but \texttt{N$_\texttt{FFT}$} might need to be increased for narrow bins, as can be seen in section $\S$\ref{sec:N5K}. The number of points in the interpolation \texttt{N$_\texttt{interp}$} captures the $k$-dependence of the growth factor and thus the default value ensures satisfying accuracy for models with mildly $k$-dependent growth factor. By default, the numerical integral over $k$ is performed over all the input values of the power spectrum, we thus define $\texttt{N}_k$ the length of the user-defined power spectrum array. Similarly to \texttt{N$_\texttt{interp}$}, $\texttt{N}_k$ has more impact at high $\ell$'s and a value of $\texttt{N}_k = 300$ is typically sufficient for mildly $k$-dependent growth factor. As mentioned before, the ratio between those two parameters plays an important role in the modelling of the $k$-dependence of the growth factor and one should aim for $\texttt{N}_\texttt{interp}/\texttt{N}_k \sim 50\%$. This is especially important for highly $k$-dependent models, in which case one might even want to further increase $\texttt{N}_\texttt{interp}$.

\newpage{}
\bibliographystyle{jcap}
\bibliography{main}

\end{document}